# Spin-Orbit Proximity Effect in Graphene

A. Avsar[1,2], J. Y. Tan[1,2], J. Balakrishnan[1,2], G. K. W. Koon[1,2,3], J. Lahiri[1,2], A. Carvalho[1,2], A. S. Rodin[4], T. Taychatanapat[1,2], E. C. T. O'Farrell[1,2], G. Eda[1,2], A. H. Castro Neto[1,2] & B. Özyilmaz[1,2,3*]

[1]Department of Physics, National University of Singapore, 117542, Singapore

[2]Graphene Research Center, National University of Singapore, 117542, Singapore

[3] NanoCore, National University of Singapore, 117576, Singapore

[4]Department of Physics, Boston University, Boston, MA, 02215, USA

*email: barbaros@nus.edu.sg

**The development of a spintronics device relies on efficient generation of spin polarized currents and their electric field controlled manipulation. While observation of exceptionally long spin relaxation lengths make graphene an intriguing material for spintronics studies, modulation of spin currents by gate field is almost impossible due to negligibly small intrinsic spin orbit coupling (SOC) of graphene. In this work, we create an artificial interface between monolayer graphene and few-layers semiconducting tungsten disulfide ($WS_2$). We show that in such devices graphene acquires a SOC as high as 17meV, three orders of magnitude higher than its intrinsic value, without modifying any of the structural properties of the graphene. Such proximity SOC leads to the spin Hall effect even at room temperature and opens the doors for spin FETs. We show that intrinsic defects in WS2 play an important role in this proximity effect and that graphene can act as a probe to detect defects in semiconducting surfaces.**

Graphene is a promising material for both fundamental spin based transport phenomena and low power consuming spin based device applications[1-2]. Among the first predictions for graphene was the topological insulator with spin polarized edge states[3]. Furthermore, room temperature (RT) ballistic charge transport[4] and gate-tunability make graphene an ideal material for spin field effect transistors (FETs)[5]. However, graphene's extremely weak intrinsic spin-orbit coupling (SOC)[6] makes the experimental observation of these phenomena very challenging. Previous studies have shown that the weak hydrogenation of graphene can significantly enhance the SOC[7-8]. However, chemical functionalization introduces disorder and severely limits charge mobility and hence, the spin relaxation length[8-9]. Equally important, the hybridization between hydrogen and carbon induces a Rashba type SOC by introducing sp3 defects thereby breaking the inversion symmetry of the lattice[10]. The decoration of the graphene surface with heavy metal adatoms such as gold (Au), indium (In) or thallium (Tl) has also been proposed to increase the spin orbit coupling[11-12]. However, also this approach introduces disorder and hence, equally limits both charge and spin transport properties[13].

These challenges have recently brought to light other 2D crystals such as transition metal dichalcogenides (TMDC). TMDCs have band gaps in a technologically attractive energy range (~1-2 eV) and orders of magnitude higher SOC than graphene[14]. This, combined with spin split bands makes them an almost ideal material for spintronics. However, the presence of unavoidable defects in TMDCs typically limits their electron mobilities to $\mu \sim 100$ cm$^2$/V.s[15]. Hence, we expect a spin relaxation length even lower than in weakly hydrogenated graphene. In this work, we show that by utilizing the proximity effect between graphene and TMDCs one can keep the extraordinary high mobilities of crystalline graphene and at the same time enhance its SOC to allow electric field controllable spin currents at room temperature.

Since graphene and WS$_2$ have comparable workfunctions (~4.6 eV and ~ 4.7 eV, respectively)[16-18], the charge neutrality point of graphene on WS$_2$ is expected to be located approximately in the middle of the energy gap of WS$_2$. Therefore, naively, one would expect WS$_2$ to act just as another inert substrate such as SiO$_2$ or BN. Remarkably, the presence of intrinsic disorder in WS$_2$ creates electronic midgap states which hybridize with the electronic states of graphene. Because of the finite thickness of the WS$_2$ (~ 10nm) substrate, the number of these defects is large. Thus, above a threshold back gate voltage ($V_{TH}$) they act as a sink for electronic charges, resulting in a gate bias ($V_{BG}$) *independent* conductivity. For $V_{BG} < V_{TH}$, our devices have electronic mobilities as high as 50.000 cm$^2$/ V.s[19], comparable to what has been reported on BN substrates[20] with no signature of SOC enhancement. In contrast, for $V_{BG} > V_{TH}$, we found a giant enhancement of SOC (~17meV). The combination of high charge mobility and large proximity induced SOC results in a gate tunable spin Hall effect (SHE) even at RT, a crucial ingredient in realizing spintronics paradigms such as the spin FET[5] and potentially even the quantum spin Hall insulator[3].

The fabrication of graphene/$WS_2$ heterostructure devices starts with the standard dry transfer method[4] of single layer graphene (SLG) onto atomically flat few layers of $WS_2$ (Fig. 1A). Subsequently, e-beam lithography is used to form electrodes and pattern graphene into a Hall bar. Typical channel lengths and widths are $l = 3\mu m$ and $w = 1\mu m$ respectively. A pair of electrodes is also created directly on $WS_2$. An optical image of a typical final device is shown in Fig. 1B. Experiments were carried out in two different measurement configurations. In the conventional local four terminal measurement configuration (Hall bar), current $I_{16}$ flows between contact 1 and contact 6 and a local voltage drop is measured between contact 2 and contact 4. In the non-local measurement configuration (H bar), the current $I_{23}$ flows between the pair of contact 2 and contact 3 and a non-local voltage $V_{45}$ is recorded across the neighboring pair of contact 4 and contact 5(Fig. 1C). All measurements were performed with standard ac lock-in technique at low frequencies in vacuum as a function of magnetic (B), back gate voltage ($V_{BG}$) at both room temperature (RT) and liquid Helium temperatures. In total, we have characterized 10 Graphene/$WS_2$ devices. Here we discuss 2 representative devices and unless otherwise stated, the results shown are from device B.

Fig. 2A shows the local conductivity σ of graphene on a $WS_2$ substrate as a function of $V_{BG}$ at T= 1.5 K ($\mu_h \approx$ 18,000 $cm^2$/V.s,). While on the hole side σ is linear in $V_{BG}$, on the electron side σ saturates above a threshold voltage $V_{TH} \geq 15V$. A similar behavior has been observed in all graphene on $WS_2$ devices and is absent e.g. in graphene on BN devices. We also performed two terminal I-V measurements directly across $WS_2$ alone. To minimize any shunting of the current through the WS$_2$, we use Cr/Au contacts (see SI). In such measurements, only an insulating response over the same V$_{BG}$ range is observed[19]. Next, we discuss perpendicular magnetic field (B$_\perp$) dependent measurements as shown in Fig. 2B-C. From Hall effect measurements at low fields, we see that the origin of the conductance saturation on the electron side is due to the saturation of the charge density at ~ 7x10$^{11}$ cm$^{-2}$ (Inset-Fig. 2C). The electron mobility above the saturation region is estimated to be $\mu_e \approx$ 24,750 cm$^2$/V.s. In high B$_\perp$ fields, this surprising result even alters the well known Landau level (LL) fan diagram of graphene. This is clearly seen in the fan diagram where the longitudinal resistance, R$_{XX}$, is recorded on a color plot as a function of both V$_{BG}$ and the perpendicularly applied B$_\perp$ (Fig. 2B). For example, at B$_\perp$ =4.5T, R$_{XX}$ vs V$_{BG}$ shows the regular integer quantum hall effect on the hole side with filling factor $v$=2, 6, 10, 14, 18, 22. However, on the electron side only the quantization for the first two LLs is observed. Since the LLs in graphene are not equally spaced in energy, it is possible to lock the graphene resistance even at the first LL for | B$_\perp$| > 10T (Fig. 2C), making such graphene/WS$_2$ heterostructure potentially a suitable device for quantum resistance metrology[21].

We now turn our attention to non-local transport measurements. This measurement technique is a powerful tool to detect spin dependent transport phenomena that otherwise escape notice. Recently, this configuration has been for example used to detect Zeeman interaction induced spin currents in graphene in the presence of an externally applied perpendicular magnetic field[22-23]. Balakrishnan *et al*, also implemented this geometry to demonstrate an enhancement of SOC in hydrogenated graphene by measuring spin precession in an in-plane magnetic field[10]. We first discuss the RT non-local measurement of a graphene on SiO$_2$ device of comparable mobility ($\mu \approx$ 13.000 cm$^2$/V.s) (Fig. 3A). For such devices the non-local resistance is *symmetric* around the Dirac peak and its magnitude over the entire back gate bias range can be fully accounted for by an Ohmic leakage contribution[24], $\sim \rho e^{-\pi L/w}$ (dashed dotted fit to the data). In contrast, graphene on $WS_2$ substrates (Devices B and C, Fig. 3B and C) shows a non-local signal which cannot be accounted for by an Ohmic contribution neither quantitatively nor qualitatively. First, the magnitude of the observed non-local signal is almost an order of magnitude larger than in graphene on $SiO_2$ samples. More importantly, we observe a strong electron-hole *asymmetry* in non-local resistance, which is completely absent in both graphene on $SiO_2$ and even in graphene on BN substrates (not shown). For instance, device B and device C have a non-local signal difference $\Delta R$ between the hole side and electron side of $\Delta R \sim 8\Omega$ and $\Delta R \sim 20\Omega$, respectively. We find that there is a sample to sample variation of the non-local signal[19] most likely due to a variation in the amount of the disorder in each sample. However, the electron-hole asymmetry in the non-local signal is observed in all our graphene-WS2 devices. Equally important, the onset of the asymmetry coincides with $V_{TH}$ at which the local conductivity saturates. Note that this asymmetry does rule out thermal effects as discussed in ref.23 as the dominant effect.

To determine the origin of the non-local signal, we applied a magnetic field $B_\parallel$ in the plane of the device parallel to the current flow direction. Such Hanle spin precession measurements for device B as a function of $V_{BG}$ are summarized in Fig. 4D. Also here we see a clear electron-hole asymmetry. The critical gate voltage at which the field dependence of the non-local signal changes dramatically coincides again with $V_{TH}$ at which the conductance saturation is observed. For all $V_{BG} < V_{TH}$, the non-local signal is magnetic field *independent*. Two representative $R_{NL}$ vs $B_\parallel$ traces at $V_{BG} = -37$ V and $V_{BG} = 2$ V (DP) are shown in Fig. 3D. On the other hand For all $V_{BG} > V_{TH}$, the nonlocal signal shows a maximum at zero magnetic field and decreases with increasing $|B_\parallel|$ with the onset of an oscillatory behavior clearly visible for $|B_\parallel| > 8$ T. Remarkably, the maximum observed change of the non-local signal $\Delta R = 8\Omega$ as a function of field is comparable to the difference observed in the non-local signal $\Delta R$ as a function of $V_{BG}$ (Fig. 3B). It is important to note that in this measurement configuration only a spin current can give rise to a magnetic field dependent signal[25]. Thus, the observation of oscillatory magnetic field dependent signal directly proves the spin origin of the non-local signal. We are now ready to extract the spin transport parameters of our devices. For this we fit the field dependence of the nonlocal signal using the following equation[24]:

$$R_{NL} = \frac{1}{2}\gamma^2 w \rho \operatorname{Re}\left[\left(\sqrt{1+i\omega_B \tau_s}/\lambda_s\right) e^{-(\sqrt{1+i\omega_B \tau_s}/\lambda_s)L}\right]$$

where $\gamma$ is the spin Hall coefficient, $w$ the width of the channel, $\rho$ is the resistivity of the channel, $\omega_B$ is the Larmor frequency, $\tau_s$ is the spin relaxation time, and $\lambda_s$ is the spin relaxation length. We obtain $\lambda_s = 2$ μm, $\tau_s$ of 5 ps. The strongly reduced $\tau_s$ when compared with $\tau_s$ of even hydrogenated spin valves (9) (~ 0.7ns) is direct consequence of the colossal enhancement of SOC.

We finally calculate the proximity SOC strength $\Delta_{SO}$. Since the dominant dephasing mechanism in SLG is Elliott-Yafet type (1,26) we have $\Delta_{SO} = \varepsilon_F \sqrt{(\tau_P/\tau_S)}$ (27) where $\varepsilon_F$ is the Fermi energy and $\tau_P$ is the momentum relaxation time, $\tau_P = h\sigma/2e^2 v_F \sqrt{(\pi n)}$. Thus, we obtain a proximity SOC of 17.6 meV which is even higher than that achieved by hydrogenating graphene[8]. With this, we identify the SHE and inverse SHE as the origin of the observed non-local signal. The latter is only possible because of the three order of magnitude enhancement of the intrinsic SOC graphene when in proximity with $WS_2$. Low field magneto resistance measurements provided further evidence for a strong enhancement of the spin-orbit coupling for $V_{BG} > V_{TH}$. Here, we see at the same threshold voltage a transition from positive to negative MR.

This change from weak localization to weak-antilocalozation behavior provides independent evidence for the proximity effect induced SOC (See S.I.).

Next we discuss the origin of the proximity effect in more detail. Recently single sulphur vacancies have been observed in MoS$_2$ flakes by using an atomic resolution direct imaging technique[28-29]. Similar structural defects are also expected for WS$_2$ crystals. For this purpose, we characterized our CVD grown WS$_2$ crystals with X-ray photoelectron spectroscopy (XPS). Fig. 4A shows the XPS spectra of W$_{4f}$ and S$_{2p}$ from where we estimate a large concentration of sulphur vacancies of the order of $10^{13}$cm$^{-2}$ (See S.I. for discussion). This result shows that there are localized states in the gap of WS$_2$, originating from sulphur vacancies. As a next step, we performed ab initio fully relativistic density functional theory (DFT) calculations to discuss the relevance of these states on the observed phenomena. Our calculations for the band structure of the graphene/WS2 interface show that the Dirac point of graphene is within the WS$_2$ band gap (Fig. 4B). The band structure of the interface is nearly a superposition of the band structures of the two moieties. The lowest conduction band of WS2 is practically unchanged by the presence of graphene, and its position relative to the Fermi level differ less than 0.1eV from those obtained by the alignment of the vacuum level for the two independent systems. Note that the proximity of WS$_2$ without vacancies does not induce any significant SOC. A sulphur vacancy (V$_S$) on the other hand introduces a double unoccupied state in the top half of the band gap of bulk WS$_2$ (Fig. 4C), as close as ~ 0.1eV to the Fermi level of graphene. In fact, using $\varepsilon_f = \hbar v_f \sqrt{\pi n}$ where $v_f$ is the Fermi velocity (0.9x10$^6$ m /s) and $n$ is the carrier concentration (7x10$^{11}$ cm$^{-2}$) at V$_{TH}$ we obtain $\varepsilon_f$~ 0.09 eV at V$_{TH}$ in good agreement with our theoretical estimate. The mid gap states originate on the vacancy dangling bonds and are localized on W. They have an unusually large spin splitting of about $\Delta E_{SO}$ ~ 0.2eV to which we assign the origin of the proximity SOC observed. The graphene electrons move from graphene to the vacancy states in WS2 by quantum tunneling. This tunneling relies on the proximity of the wavefunction of the vacancy states to the graphene, which is characterized by hybridization energy $V_H$. From second order perturbation theory ($\Delta E_{SO} \approx V_H^2/\Delta_{SO}$), the hybridization energy is estimated to be 0.059 eV.

In conclusion, we have demonstrated that graphene is very sensitive to the defects in the underlying WS$_2$ substrate. These structural defects enhance the weak SOC of graphene while preserving the extraordinary electronic properties. Since all TMDCs have similar structures and properties, this proximity effect is expected to be present in a wide range of different crystals. Our results open a new avenue for the realization of many exotic phenomena such as the quantum SHE.

**Acknowledgments:** We thank A. Hamilton and R. Ribeiro for their help and useful discussions. This work was supported by the Singapore National Research Foundation Fellowship award (RF2008-07-R-144-000-245-281), the NRF-CRP award `Novel 2D materials with tailored properties: beyond graphene' (R-144-000-295-281) and the Singapore Millennium Foundation-NUS Research Horizons award (R-144-001-271-592; R-144-001-271-646).

**Author Contributions**

B.Ö. and A.H.C.N. devised and supervised the project. A.A. and B.Ö. designed the experiments. A.A. and J.Y.T. did the device fabrication. A.A, J.B, G.K.W.K., T.T, and E.C.T.O carried out the transport measurements. A.H.C.N., A.C. and A.S.R. developed the theoretical aspects of this work. E.G. and J.L grew WS2 crystals and performed XPS measurements. A.A., A.C., A.H.C.N.

and B.Ö. co-wrote the manuscript. All authors discussed the results and commented on the manuscript.

**Additional information**

Supplementary information is available in the online version of the paper. Reprints and permissions information is available online at www.nature.com/reprints. Correspondence and requests for materials should be addressed to B. Ö.

**Competing financial interests:**

The authors declare no competing financial interests.

**Figure Captions:**

**Fig. 1. Device Characterization:** (A) Schematics representation of a multilayer $WS_2$/Graphene heterostructure device. The highest unoccupied state of the sulphur vacancy is depicted in yellow, highlighting on the W atoms closest to the vacancy. W, S and C atoms are represented by dark gray, orange and light gray spheres, respectively. (B) Optical micrograph of a completed device with multiple Hall bar junctions on G/ $WS_2$ heterostructure and a two terminal device on $WS_2$. The scale bar is 2μm. (C) Schematics for the local and non-local measurement configurations.

**Fig. 2. Electronic transport measurement in graphene on $WS_2$ substrate:** (A) Local resistivity (black lines) and conductivity (red line) measurement as a function of back gate voltage at 1.5 K. (B) Landau fan diagram of longitudinal resistance as a function of magnetic field and back gate voltage. (C) Corresponding plots of longitudinal resistance as a function of back gate voltage at constant magnetic fields. (Black and red lines represent 4.5 T and 12 T respectively.) <u>Inset:</u> Carrier concentration as a function of applied back gate voltage.

**Fig. 3. Spin transport measurement in graphene on WS$_2$ substrate:** (A) Nonlocal resistance measurement as a function of back gate voltage at RT in a reference graphene/SiO$_2$ device, Sample A, (red line) with its calculated Ohmic contribution (black line). (B-C) Nonlocal measurement for Sample B and Sample C at 1.5K and RT respectively. (Fig.3. B is the focused area of Inset Fig.3. B) (D) Fan diagram of nonlocal resistance as a function of in-plane magnetic field and back gate voltage. Color scale bar is adjusted to show between 12-20 Ω for clarification. Corresponding plots of non-local resistance as a function of in plane magnetic field at constant back gate voltages.(Black, pink and red lines represent 37 V , 2V (DP) and 37 V respectively.)

**Fig. 4. XPS measurement of WS$_2$ and DFT bandtructure calculations:** (A) High resolution XPS of W$_{4f}$ and S$_{2p}$ core levels. (B) Bandstructures for the interface between graphene and monolayer WS$_2$ and sulphur vacancy in bulk WS$_2$. Since the local or semi-local approximations for the exchange functional are known to result in an underestimation of the bandgap (our estimated value is 0.2eV smaller than the experimental value in ref. (30)), the unoccupied electron states in latter are shifted 0.2eV to correct the bandgap.

# Figures

**Spin-Orbit Proximity Effect in Graphene**

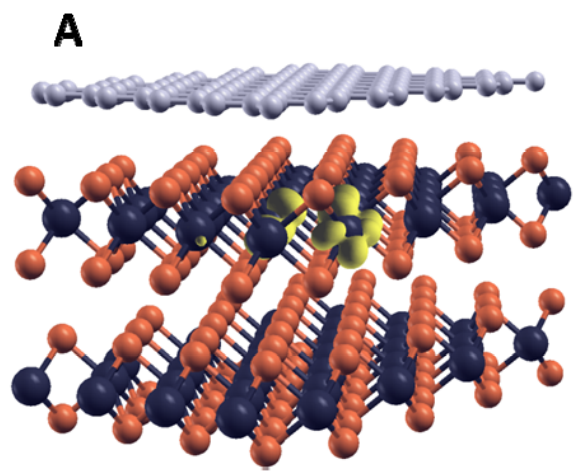 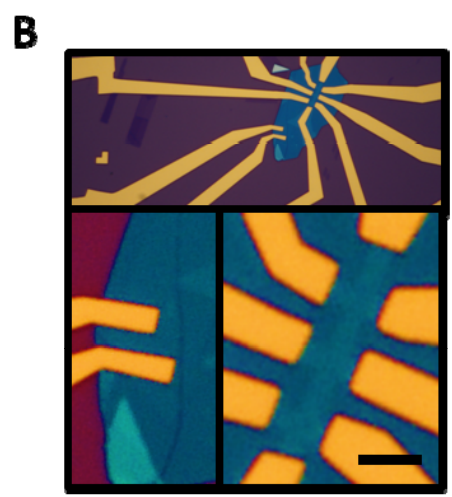 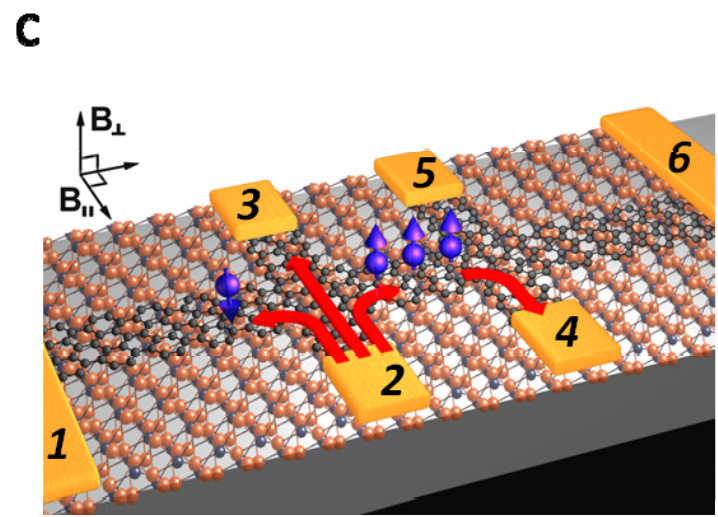

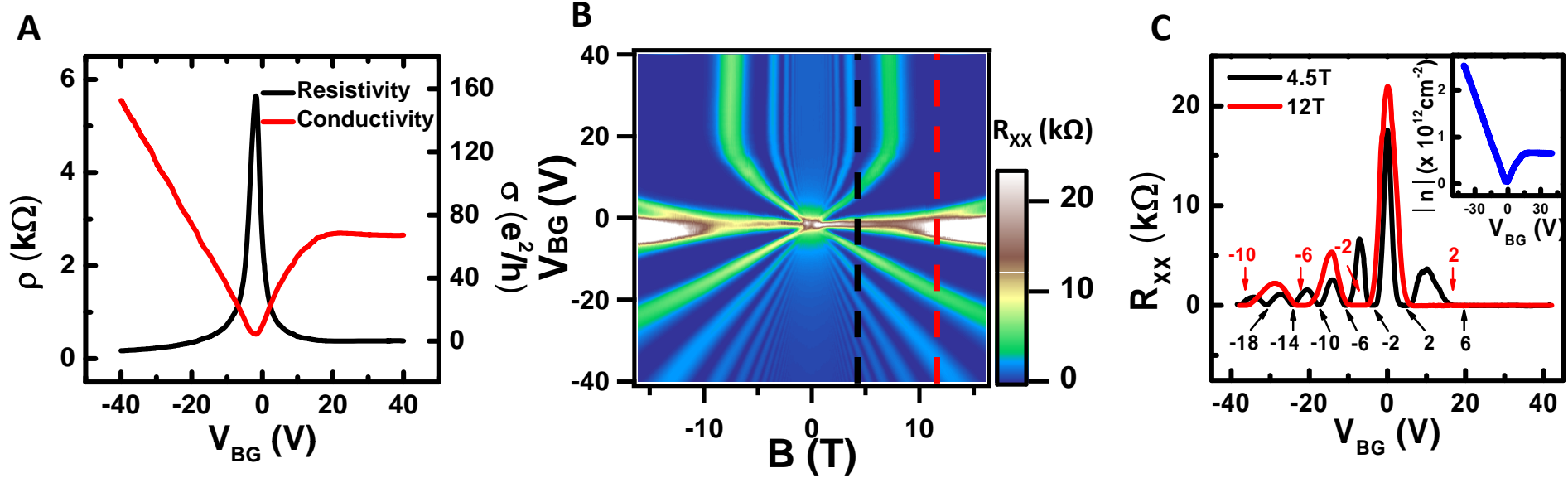

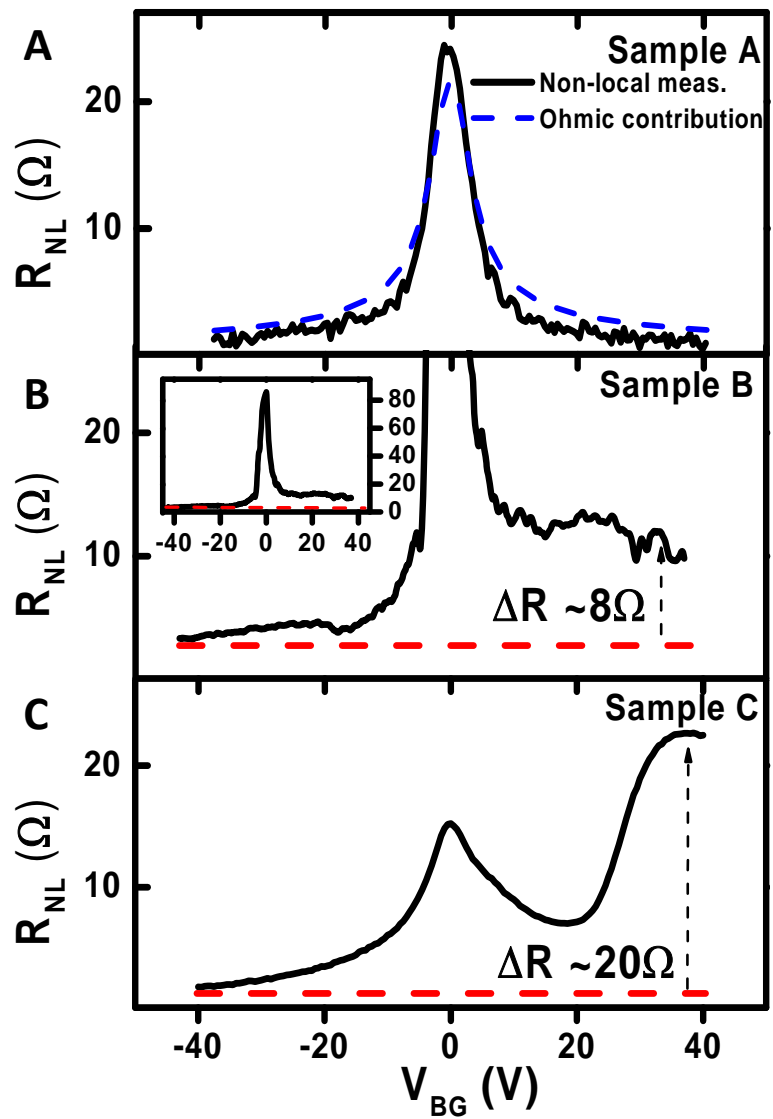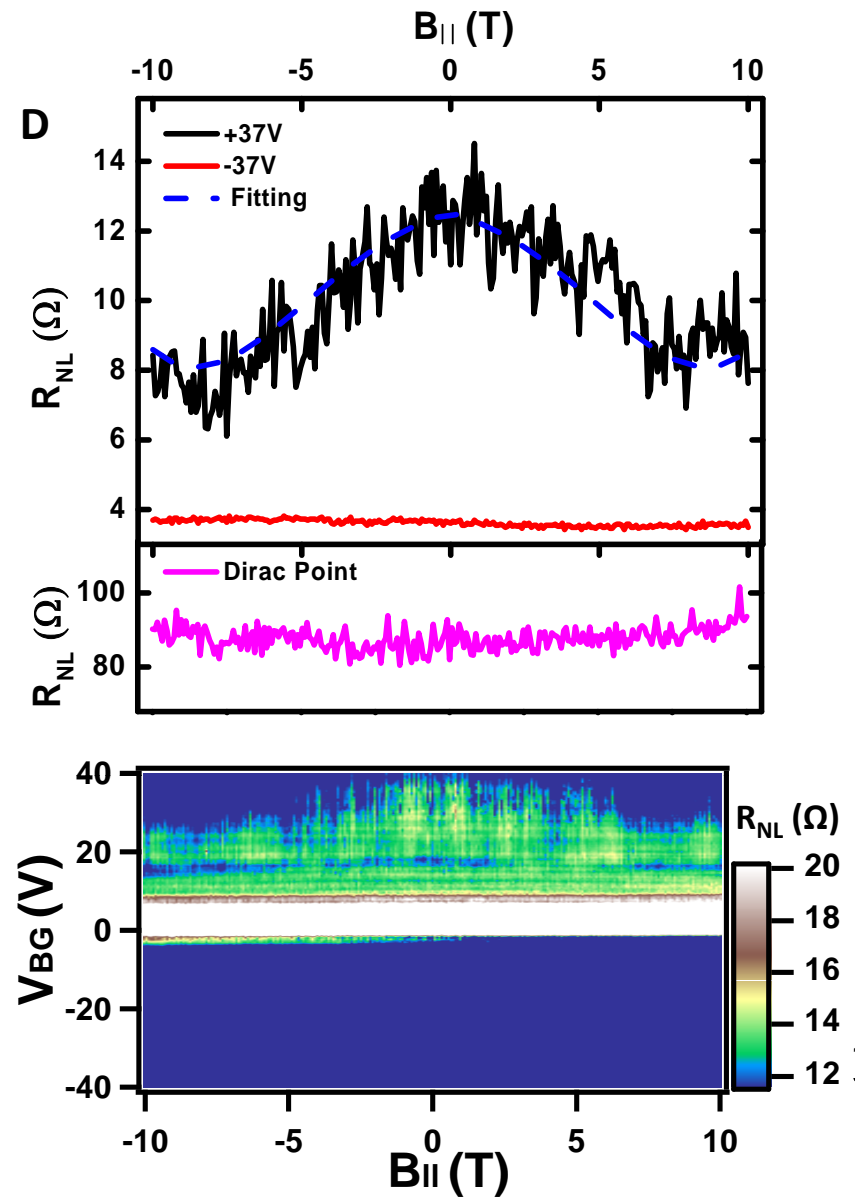

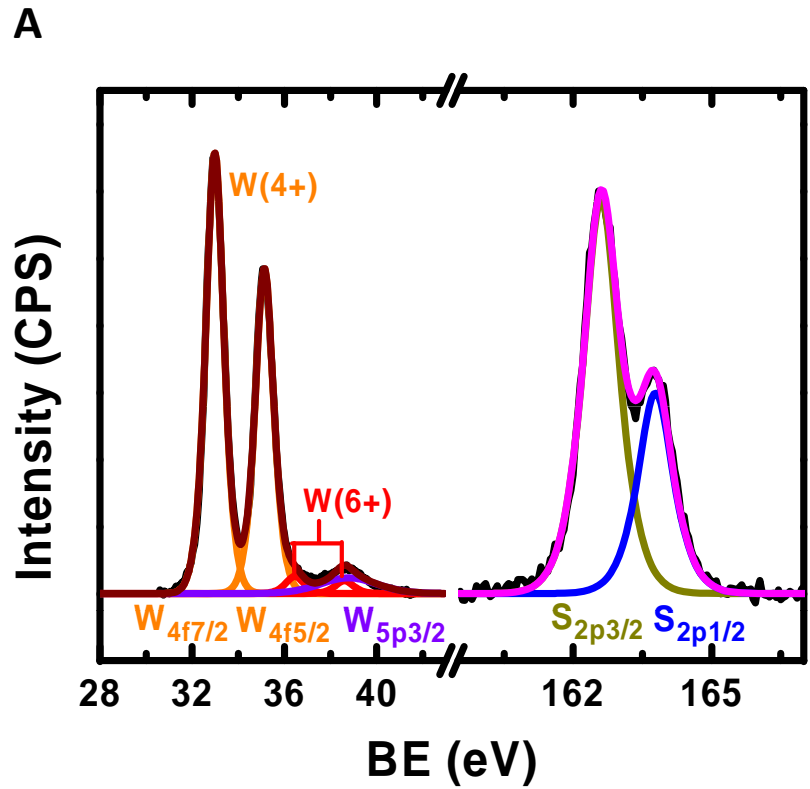 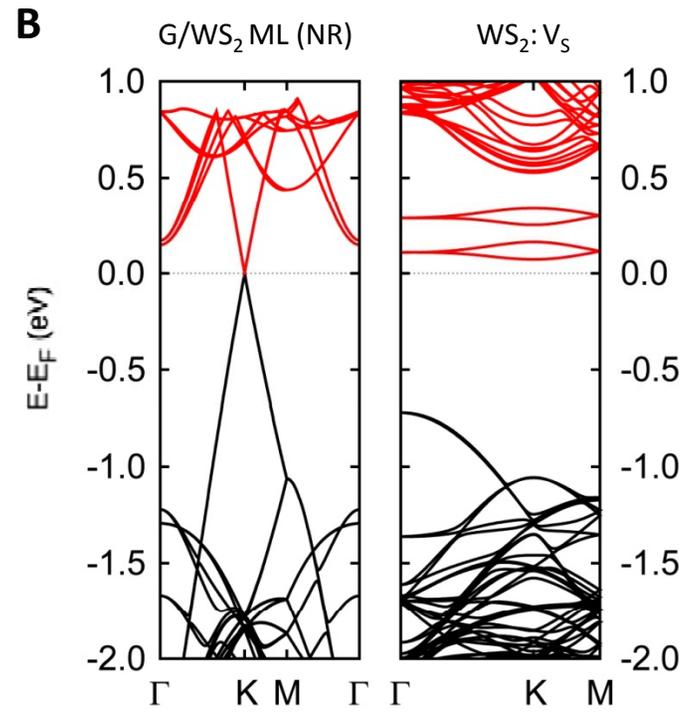

# Supplementary Materials for

## Spin-Orbit Proximity Effect in Graphene

A. Avsar[1,2], J. Y. Tan[1,2], J. Balakrishnan[1,2], G. K. W. Koon[1,2,3], J. Lahiri[1,2], A. Carvalho[1,2], A. S. Rodin[4], T. Taychatanapat[1,2], E. C. T. O'Farrell[1,2], G. Eda[1,2], A. H. Castro Neto[1,2] & B. Özyilmaz[1,2,3]*

[1]Department of Physics, National University of Singapore, 117542, Singapore

[2]Graphene Research Center, National University of Singapore, 117542, Singapore

[3] NanoCore, National University of Singapore, 117576, Singapore

[4]Department of Physics, Boston University, Boston, MA, 02215, USA

*email: barbaros@nus.edu.sg

**This PDF file includes:**

Supplementary Text

Figs. S1 to S10

References

## 1. AFM and Raman characterization of $WS_2$ crystals

$WS_2$ flakes are prepared on $SiO_2$ substrates and then annealed in a gaseous mixture of Ar / $H_2$ (9/1) at 350°C for 3 hours to ensure the surface is free of residues. Figure S1a shows the typical optical and AFM images of few layers $WS_2$ with 12 nm thickness after the annealing process. The thickness of $WS_2$ substrates used in this study varied from 5 nm to 15 nm. Device B and Device C discussed in the main text have $WS_2$ thickness of 9 nm and 7 nm respectively. The surface topography shows that the surface roughness in $WS_2$ (~110 pm) is comparable to that of Boron Nitride[1].

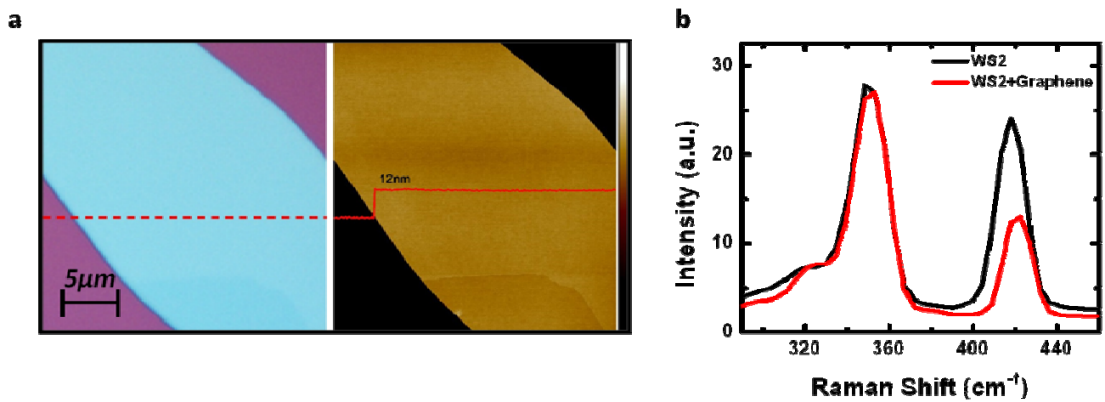

Figure S1. Crystal characterization. **a -** Optical and AFM images of a representative $WS_2$ flake. Color scale of the AFM image represents 0-20 nm. **b -** Raman spectrum of few layers $WS_2$ with and without graphene.

Figure S1b shows the typical Raman spectra of few layers $WS_2$ before and after graphene transfer. An excitation wavelength of 532 nm is used. The Raman spectrum is dominated by two peaks at 351 $cm^{-1}$ and 418 $cm^{-1}$ similar to those reported in reference 2. The peak at 351 $cm^{-1}$ is the overlapping of a first-order mode ($E^1_{2g}$ ($\Gamma$)) and second order mode that is activated by disorder (2LA (M)). The peak at 418 $cm^{-1}$ is due to another first order mode at the Brillouin zone, $A_{1g}$ ($\Gamma$). Unlike the peak at 351 $cm^{-1}$, the intensity of the peak at 418 $cm^{-1}$ is reduced slightly after the graphene transfer. The laser power in this measurement is kept below 0.2mW to avoid any damage to the flakes.

## 2. Growth and XPS of WS$_2$ crystal

Bulk crystals of 2H-WS$_2$ were grown by the chemical vapor transport (CVT) method using iodine as the transport agent. As-received WS$_2$ powder (Alfa Aesar, 99.9 purity %) was sealed in an ampoule of fused quartz. The ampoule was kept in a temperature gradient from 875 to 935 C in a dual zone furnace for 1 week. WS$_2$ crystals grew on the cold end of the ampoule.

X-ray photoelectron spectroscopy (XPS) measurements were performed by using non monochromatic Mg $Ka$ radiation (1253.6eV). Figure S2 shows the survey scan of a WS$_2$ crystal. Along with W and S core levels, O$_{1s}$ and C$_{1s}$ peaks are also observed. Next, we focus on the XPS spectra of W$_{4f}$ and S$_{2p}$ core levels (Fig. 4A). The W$_{4f}$ core level peak can be fitted to two sets of doublets corresponding to W (+6) and W (4+) oxidation states. The presence of W (+6) oxidation states indicates that tungsten disulphide has undergone oxidation. The observed peak position for W$_{4f7/2}$ for the +6 and +4 oxidations states matches with those reported in the literature[3]. The S$_{2p}$ peak could be fitted to only one set of doublet corresponding to S-W bonding.

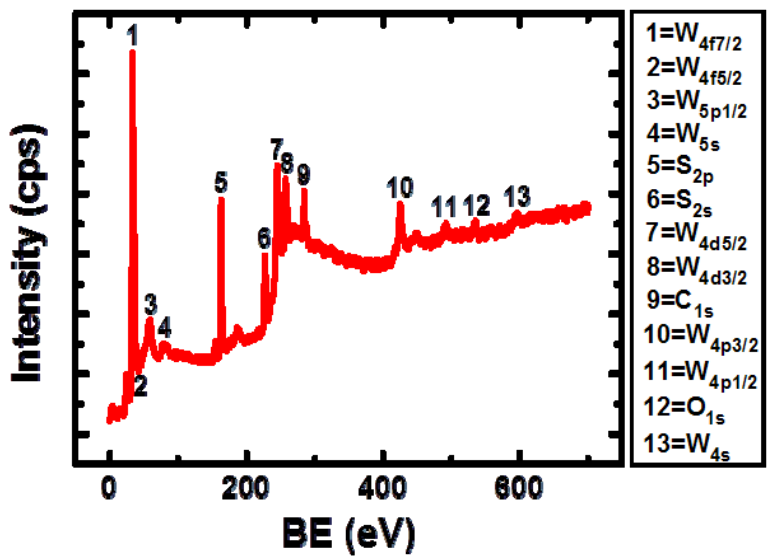

Figure S2. XPS survey scan of WS$_2$ crystal acquired with Mg $Ka$ line.

The stoichiometry of WS$_2$ crystal was obtained from:

$$\frac{[W]}{[S]} = \frac{\sigma_{S2p}(h\nu)}{\sigma_{W4f}(h\nu)} \times \frac{\lambda_{S2p}}{\lambda_{W4f}} \times \frac{I_{W4f}}{I_{S2p}}$$

where $\sigma_{S2p}(h\nu)$ and $\sigma_{W4f}(h\nu)$ are photo-ionization cross sections of the 2p and 4f core level of sulphur and tungsten, respectively. The values of $\sigma_{S2p}(h\nu)$ and $\sigma_{W4f}(h\nu)$ are 0.045 and 0.18 Mb, respectively, obtained from tabulated data[4]. $\lambda_{S2p}$ and $\lambda_{W4f}$ are inelastic mean free paths (IMFP) of the photoelectrons with kinetic energies that correspond to the S and W core levels, respectively. The values of $\lambda_{S2p}$ and $\lambda_{W4f}$ were estimated to be 1.29 nm and 1.36 nm respectively using the Seah and Dench method[5]. $I_{S2p}$ and $I_{W4f}$ are the integrated intensities of the photoelectron peaks of the S$_{2p}$ and W4$_f$ levels after fitting, respectively. The error arising from the fitting is ~4%. With these values, the [W]/[S] ratio was estimated to be 0.57-/+0.02, which corresponds to 12 % sulphur vacancies in the crystal. Thus, the key findings of the XPS studies are that (1) the WS2 crystals are notably sulphur deficient and (2) the sample is free of metal impurities in detectable concentrations. The origin of sulphur deficiency is partly due to the suspected partial oxidation and polycrystallinity of the sample. However, the estimate of 12 % is an upper limit to the sulphur vacancies for the following reasons:

First, our CVD-grown WS$_2$ crystal consists of 50~500 μm size polycrystalline grains, much smaller than the XPS probe size (~ 1 mm). The suspected presence of W terminated edge atoms (i.e. locally sub-stoichiometric) will lead undoubtedly to some errors in the estimation of the sulfur vacancies. It should also be noted that the surface terraces are expected to exhibit high concentration of edge defects. Therefore, in small scale atomically thin cleaved single crystal, WS$_2$ flakes used for our transport studies, the vacancy concentration is almost certainly smaller. The possible presence of divacancies is also like to further reduce this lower bound[6].

With regards to oxidation, we observe a small fraction of W$^{6+}$ states along with W$^{4+}$ that are expected for ideal WS$_2$. Based on our fits, this amounts to approximately 4% of the total W atoms. It is conceivable that some of the sulfur vacancies are passivated by O atoms, or conversely, that some of the W signals are coming from WO$_3$ that may be present in the sample. In either case, this leads to a lower concentration of sulfur sites that are actually vacant, setting a lower bound to around 5 % (~1.2x 10$^{14}$ cm$^{-2}$).

Note also that W. Zhou et al.[6] and H. Wiu et al.[7] and have recently provided direct evidence that sulphur vacancies are abounded in CVT grown molybdenum disulfide. Since TMDCs have similar structures and properties, similar sulphur vacancy concentration in $WS_2$ is expected. Both results, ours based on XPS and theirs based on TEM imaging is in order of magnitude agreement for.

Based on all these considerations, we conclude even just assuming the lower bound it is reasonable there is a large concentration of sulphur vacancies of the order of $10^{13} cm^{-2}$. This is significantly larger than the concentration of other impurities and thus the vacancies play a dominant role in the phenomenon that is reported in this study.

## 3. First-principle calculations

Our XPS measurements clearly indicate that $WS_2$ crystals have oxygen impurities and sulphur deficiency. To check the effect of such defects we have performed non-relativistic DFT calculations by using the method described in reference 8. For the dispersion we used the correction by Grimme[9]. Bulk $WS_2$ was modelled using a 4×4 × 1 supercell, with the experimental c/a ratio. The graphene-WS2 monolayer interface was modelled in a supercell with 3×3 primitive cells of $WS_2$ and 4×4 primitive cells of graphene, with the respective lattice vectors aligned.

We investigated Si and O replacing for S (noted $Si_S$ and $O_S$), adsorbed oxygen (noted $O_{ad}$) and S and W vacancies. The substitutional configuration is shown in Fig. S3-a. The two configurations found by Ataca et al. [S10] are depicted in Fig. S3. b-c. The $O_{ad}$ configuration II ($Mo_1$ in Ref. 10) is 1.7eV higher in energy than the $O_{as}$ configuration I ($S_4$ in Ref. 10). Finally, the S and W vacancies are modelled by removing the respective atoms. In both cases there is little lattice reconstruction.

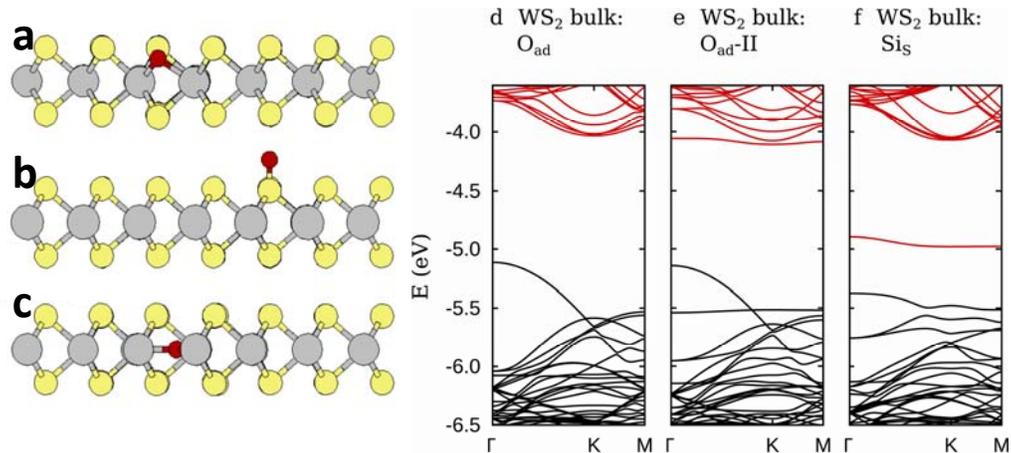

Figure S3. Geometry of a- O substituting for S b- adsorbed oxygen on the S layer c- adsorbed oxygen in the W layer. Bandstructures of defects in bulk $WS_2$: oxygen in the d,e - $S_4$ and $Mo_1$ positions, and f- substitutional silicon.

The non-relativistic bandstructure of bulk supercells containing defects with levels in the gap are shown in Fig. S3d-f. The position of the conduction band is underestimated by 0.2 eV within the working approximations. This underestimation is systematic and well known.

**Stability**

Since the only oxygen defect that is an electron acceptor is metastable, we investigated how likely it is to be present in the samples in that configuration. As the local bonding environment and even the bandstructure of the defects considered here are very similar in bulk and monolayer $WS_2$, we used a monolayer to model their stability:

1. Above the temperature of mobility of the sulphur vacancy: In thermal equilibrium, the stability of a substitutional atom $X_S$ versus an adsorbed atom of the same species $X_{ad}$ can be characterized by the enthalpy change associated with the capture of the impurity atom a sulphur vacancy: $V_S+X_{ad} > X_S$ where Vs is the sulphur vacancy and $X_{ad}$ is the adatom. For oxygen, this reaction enthalpy is -3.5 eV and -5.2 eV in configurations I and II, respectively. For silicon, the reaction enthalpy is -2.8 eV. These values are larger than the formation energy of a sulphur vacancy in a S-poor material. Therefore, Si and O occupy predominantly substitutional positions.

2. Below the temperature of mobility of the sulphur vacancy: We also investigated the hypothesis that, at temperatures at which Vs is immobile, oxygen can be introduced into the metastable position II. The calculated activation energy for the transformation $O_{ad}$-I -> $O_{ad}$-II is 4.3 eV in the forward direction. This energy barrier prevents the introduction of $O_{ad}$ into the II site below 590K.

## 4. Charge transport in few layer $WS_2$

In both local and non-local measurements in graphene on $WS_2$ substrate, a 100 nA a.c. current is used while the back gate voltage ($V_{BG}$) is swept from 40V to -40V. Devices are fabricated with Cr/Au contacts with thickness of 2 nm/ 100 nm.

It is well known that it is challenging to form Ohmic contacts to semiconducting 2D crystals[11-12]. In order to minimize the charge injection into $WS_2$ in our heterostructure devices, we deliberately look into a metal that exhibit large contact resistance. A recent study by McDonnell et al. reveals that Cr reacts with TMCS leading to formation of a highly defective and resistive interface[13]. Unlike Cr, Ti is reported not to react with MoS2. M. Fontane et al. also reported that MoS$_2$ is very strongly n-type dope if contacted with Cr/Au (2nm/100nm) such that no conduction is observed until $V_{BG}$ > 60V[14]. We observe a similar behavior on $WS_2$. Fig. S4 compares the transport characteristic of few layers of $WS_2$ contacted with Cr/Au (2 nm/100 nm) and Ti/Au (2nm /100nm) contacts. The WS2 sample with Cr/Au shows an insulating response under a bias voltage of 0.1V with a gate sweep from 40V to -40V. On the other hand, the WS2 sample fabricated with Ti/Au contacts shows a typical semiconducting behavior. Thus, we choose Cr/Au contacts for our graphene/WS2 heterostructure devices so that we make Ohmic contact to graphene but not to $WS_2$ directly. This minimizes the chance of having a small but finite, parallel conducting channel.

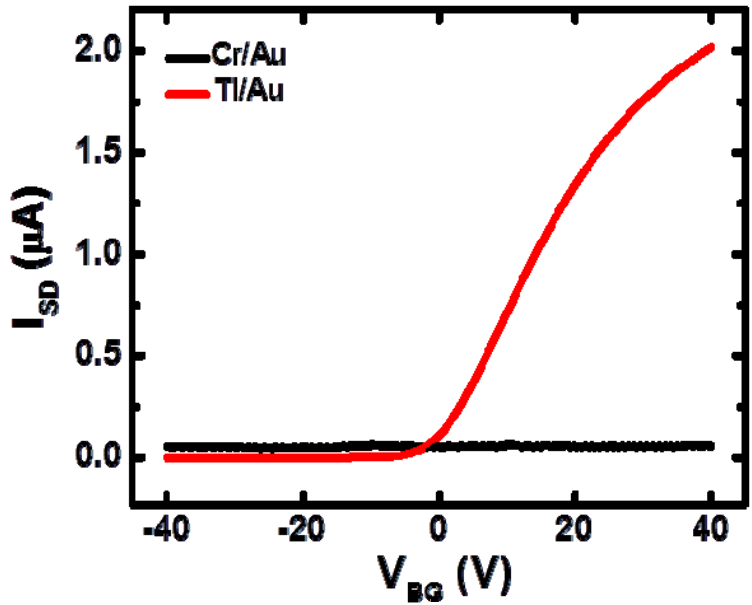

Figure S4. Two terminal resistance measurement in few layers of $WS_2$ flakes fabricated with Cr/Au and Ti/Au contacts. With Cr/Au contacts the source drain current remains negligibly small over the entire gate voltage range ($I_{SD}$<50nA).

## 5. Charge transport in Graphene/$WS_2$ heterostructure

Figure S5 shows the resistivity and conductivity of graphene on $WS_2$ substrate (Sample D) as a function of $V_{BG}$ at 6K. Also, for this sample the conductivity is linear in $V_{BG}$ on hole side, but sample exhibits a $V_{BG}$ independent conductivity above a threshold back gate voltage, ($V_{TH}$ ~15V) . The sample has a field effect mobility of ~ 50.000 $cm^2$/V.s at low densities is extracted using              , similarly to the reported values for graphene on BN substrates[1].

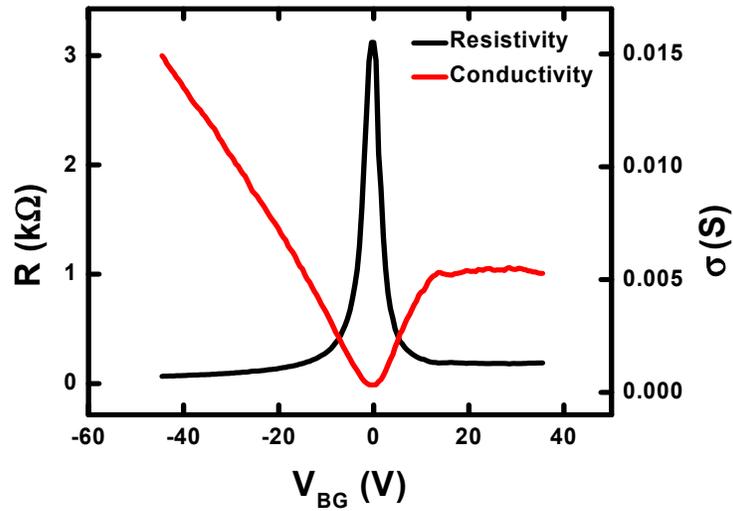

Figure S5. Resistivity and conductivity plots (black and red curves respectively) as a function of backgate voltage in graphene on WS$_2$ substrate at 6K.

## 6. Charge transport in inverse structure

As discussed in supplementary section #4, special attention is given to minimize the shunting of current through WS$_2$ during the transport measurements in heterostructure devices. In order to further rule out that the observed gate voltage dependent saturation in signal arises due to a gate voltage dependent conduction in WS$_2$, a top gated graphene/WS$_2$ heterostructure is fabricated by utilizing PVDF as dielectric. PVDF has been demonstrated to induce high charge carrier densities up to $3\times10^{13} cm^{-2}$ into graphene[15].

Inset Fig.S6 shows the optical image of the top gated device. Thick PVDF dielectric (500nm) is spin coated to minimize the leakage current. Figure S6 shows the conductivity of graphene on WS$_2$ substrate as a function of top gate voltage ($V_{TG}$). Similarly back gated devices; we observe the saturation in conductivity at electron side. This observation demonstrates: 1-) The saturation in graphene conductivity is not due to parallel conductance of WS$_2$ as the charge carriers in graphene is tuned through PVDF dielectric rather than WS$_2$ itself. 2-) The persistence of saturation in graphene conductance even at high $V_{TG}$ show that the density of sulphur vacancies to be at least in the order of $10^{13} cm^{-2}$. This concentration of vacancies has an order of magnitude agreement with the sulphur vacancy concentration estimated from our XPS analysis.

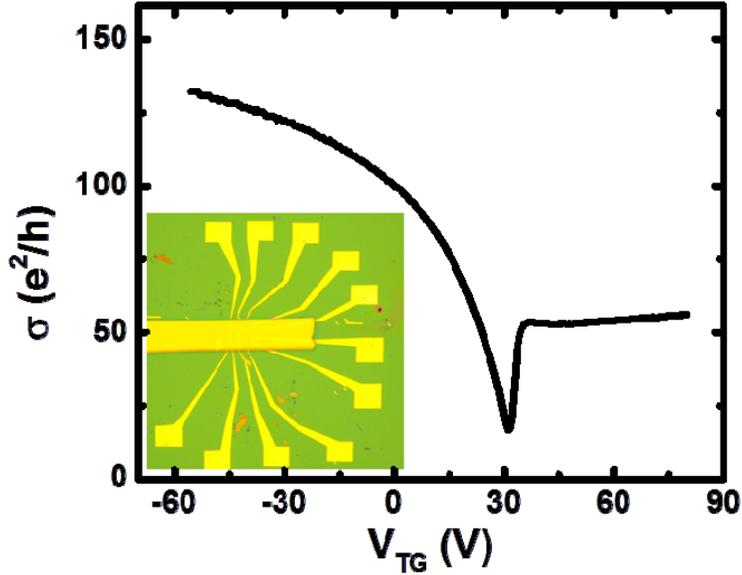

Figure S6. Conductivity of graphene as a function of top-gate voltage in "Top-gate/PVDF/Graphene/WS$_2$/SiO$_2$/Si" heterostructure at room temperature. The inset is the optical image of the measured device.

## 7. Quantum interference measurements in graphene/WS2 heterostructure

In this section we show local magneto-conductance measurements in perpendicular field, which demonstrate weak-antilocalization, providing independent evidence for strong spin-orbit coupling in the graphene-WS$_2$ heterostructure.

Fig. S7a shows conductivity against $V_{BG}$ of a graphene-WS$_2$ heterostructure as described in the main text, Fig S7b shows local conductivity against perpendicular magnetic field; at $V_{BG}<V_{TH}$ we observe positive magneto-conductance while at $V_{BG}>V_{TH}$ we observe negative magneto-conductance, as is characteristic of a crossover from weak localisation to weak antilocalization on reaching the conductivity saturation region.

In order to fit to the Maekawa-Fukuyama expression for weak-antilocalization we find it necessary to rescale the data[16]. By doing so we obtain for the spin-orbit field $H_{SO} = 0.02\pm0.02$ T at $V_{BG} = -29$ V, and $H_{SO} = 0.16\pm0.02$ T at $V_{BG} = 14$ V. These values are consistent with those obtained in, respectively, the conduction and valence band of WSe$_2$, which has extremely similar electronic structure as $WS_2$[17].

We therefore attribute this magneto conductance to quantum interference effects in the $WS_2$ substrate and suggest that the rescaling necessary to fit to the Maekawa-Fukuyama expression is due to the presence of the graphene channel, which dominates the conductivity. This effect is described in detail in a forthcoming article[18].

With regard to the mechanism for proximity induced spin-orbit coupling described in the main text, these results show that applying a positive backgate voltage, and thereby raising graphene towards the weakly spin-orbit coupled conduction band of $WS_2$, instead accesses states that have strong spin-orbit coupling which is characteristic of the valence band of $WS_2$[19]. We attribute these strongly spin-orbit coupled states within the band gap of $WS_2$ to sulphur vacancies consistent with density functional theory.

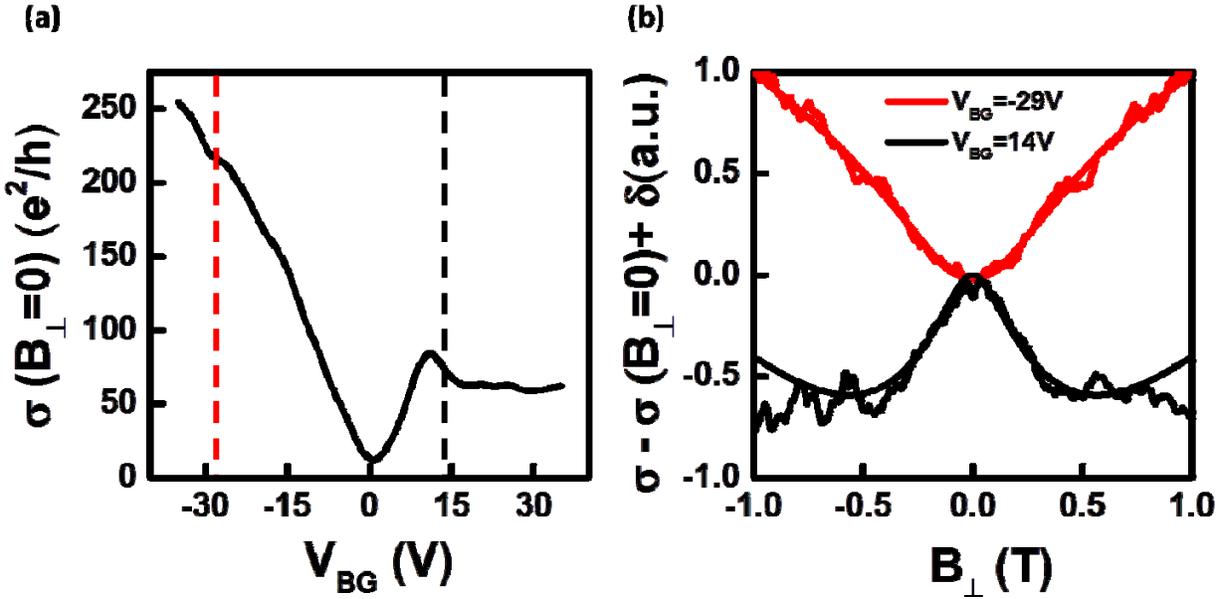

**Figure S7: (a) Conductivity of** graphene-$WS_2$ at zero magnetic field **(b)** Magneto-conductance of graphene-$WS_2$ measured at T=1.2 K in perpendicular magnetic field. The magneto conductance is rescaled and shifted to fit to the Maekawa-Fukuyama formula, the measurements show weak localization like behavior at negative backgate and weak antilocalization, associated with strong spin orbit splitting, at $V_{BG}>V_{TH}$. The black and red dashed lines in back gate voltage dependence of graphene conductivity plot shown in (a) represents the fixed back gate voltages where low field magnetic field dependent measurement was performed.

## 8. Magnetotransport measurements in high fields in Graphene/WS$_2$ heterostructure

As discussed in main text, electron carriers in graphene are populated into localized midgap states of WS$_2$ once the Fermi level aligns to these empty states. This leads to a graphene resistance that is independent of $V_{BG}$. In this section, we discuss the magnetic field ($B_\perp$) dependent resistance measurements in graphene/WS$_2$ heterostructure devices in more detail. Fig. S8a-b shows the longitudinal ($R_{XX}$) and Hall ($R_{XY}$) resistance of graphene as a function of $V_{BG}$ at fixed perpendicular to the plane magnetic fields. The hole side shows regular integer quantum hall effect that is observed in high quality graphene devices. Uniquely to this system, on the electron side both $R_{XX}$ and $R_{XY}$ saturates above a $V_{TH} \sim 15V$. By tuning the Fermi level with $B_\perp$, it is possible to keep the Landau Levels completely filled over an extraordinarily broad range of back gate voltages. The observed phenomenon allows for sustaining a nondissipative current over a large back gate range, resulting in an unusual robust quantization of the resistance.

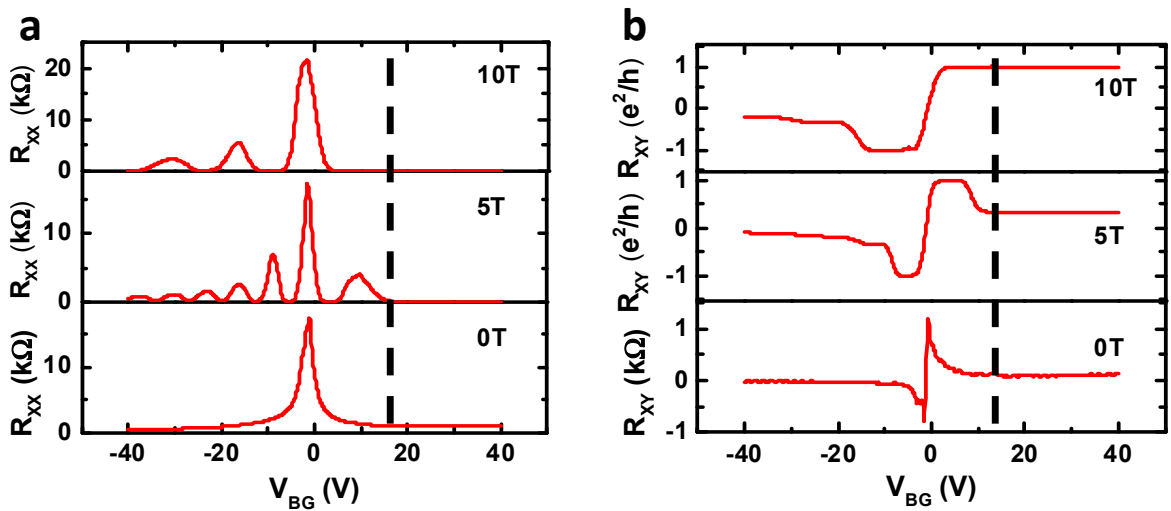

Figure S8. a,b - Longitudinal and Hall resistance measurements as a function of back gate voltage at fixed magnetic fields at T = 1.5K.

## 9. Effective mass calculation of graphene on WS$_2$ substrate

In this section, we determine the effective mass of charge carriers in graphene on WS$_2$ substrate by measuring the resistivity of sample B as a function of back gate voltage, perpendicularly applied magnetic field and temperature. Fig. S9a-b shows such Landau fan diagrams of the longitudinal resistance at 15K and 30K. The amplitude of SdH oscillations is extracted from these color plots at fixed $V_{BG}$ ranges with a step of $\Delta V=10$. Fig. S9c shows such plot at $V_{BG} = -20V$. The amplitude of SdH oscillations can be described by the following expression to extract effective mass of charge carriers: where $T$ is the temperature, is the effective mass and $h$ is the Planck constant[20]. The gate dependence of and carrier concentration is shown in Fig. S9d. While increases slightly in $V_{BG}$ on the hole side, similar to the previous report, it saturates above a threshold voltage on electron side[21].

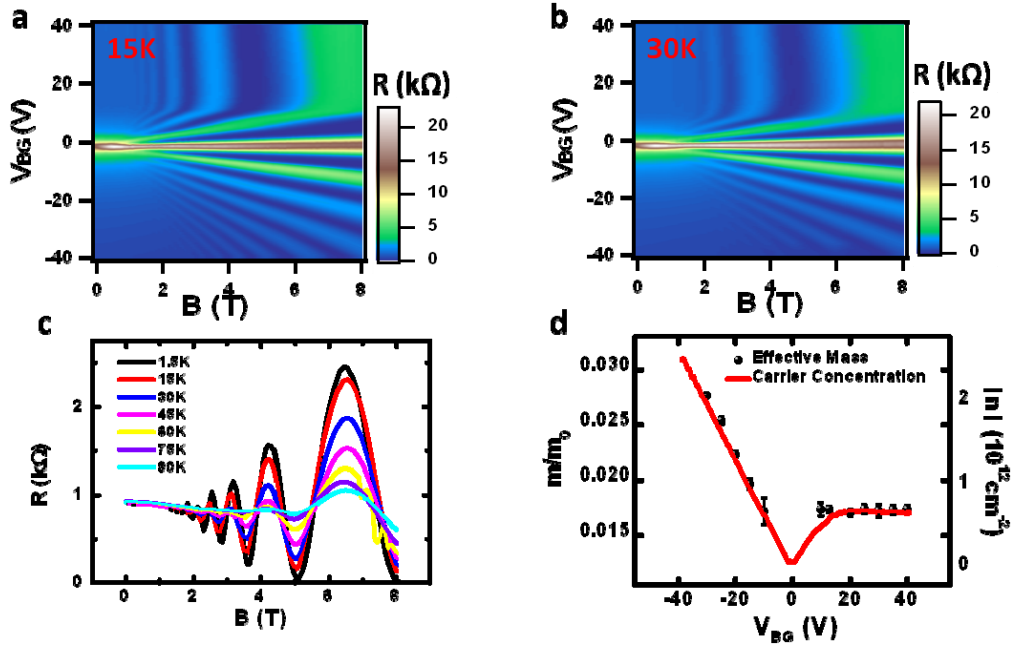

Figure S9. a,b - Landau fan plots of longitudinal resistance at 15K and 30K, c - Amplitude of SdH oscillation as a function magnetic field at different temperature values. d - Calculated effective mass and carrier concentration as a function of back gate voltage.

## 10. Additional non-local data

While the overall behavior of the non-local resistance as a function of $V_{BG}$ is similar in all samples characterized, and shows a non-zero amplitude for most samples even on the hole side. However, as discussed in main text, the spin contribution to $R_{NL}$ can be easily estimated in our devices, since the observed proximity effect exhibits a strong electron hole asymmetry. As expected, in samples with higher mobility this non-spin based background signal is much reduced. An example for such a device with a comparatively highly charge mobility ($\mu \sim 40,000 \text{cm}^2/\text{V.s}$) is shown Fig. S10. Here the black line represents are fits of the Ohmic contribution. For $V_{BG} < V_{TH}$ the non-local signal is fully accounted for by the later. For $V_{BG} > V_{TH}$ we again see a strong enhancement of the non-local signal. We attribute the fluctuations in the non-local signal near the Dirac Point to the presence of charge puddles[22].

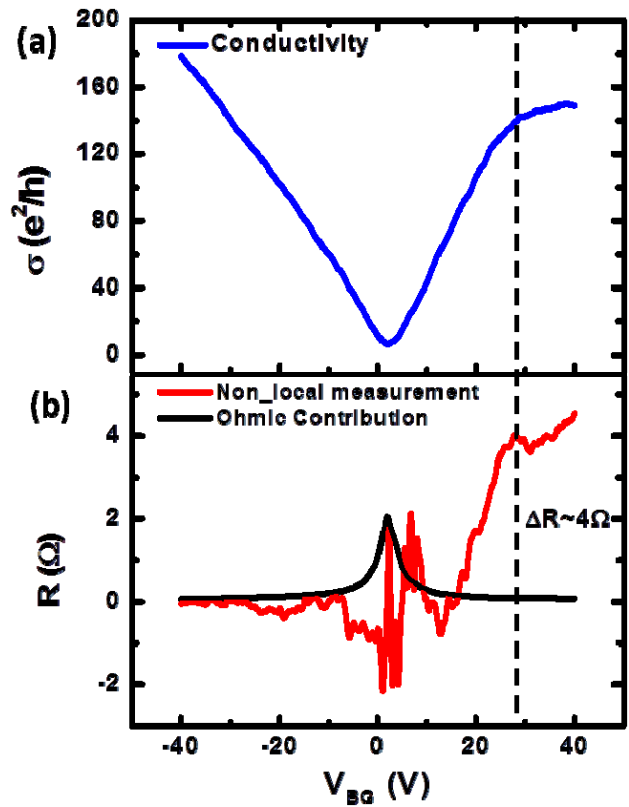

Figure S10. a,b - Local conductivity and non-local resistance measurement of graphene on $WS_2$ substrate. Black curve in b - represents the Ohmic contribution to the non-local signal. $V_{TH}$ for this sample is at 29V.